\documentclass[preprint,showpacs,preprintnumbers,amsmath,amssymb]{revtex4}
\usepackage{graphicx}
\usepackage{dcolumn}
\usepackage{bm}
\usepackage{amssymb}   

\font\scripti=cmmi7
\font\scriptscripti=cmmi5
\def\sib#1{\setbox0 = \hbox{\scripti #1}
  \kern-.02em\copy0\kern-\wd0
  \kern.04em\box0} 
\def\ssib#1{\setbox0 = \hbox{\scriptscripti #1}
  \kern-.02em\copy0\kern-\wd0
  \kern.04em\box0} 
\font\tenib=cmmib10 
\skewchar\tenib='177 \skewchar\tenib='177 \skewchar\tenib='177
\def\pbold#1{\setbox0 = \hbox{$ #1 $}
  \kern-.022em\copy0\kern-\wd0
  \kern.011em\copy0\kern-\wd0
  \kern.011em\copy0\kern-\wd0
  \kern.011em\copy0\kern-\wd0
  \kern.011em\box0} 

\def\op{\Delta}

\def\up{\uparrow}
\def\dwn{\downarrow}

\def\lesssim{\ \raise.3ex\hbox{$<$}\kern-0.8em\lower.7ex\hbox{$\sim$}\ }
\def\gesim{\ \raise.3ex\hbox{$>$}\kern-0.8em\lower.7ex\hbox{$\sim$}\ }

\begin{document}
\title{Strong-coupling corrections to ground-state properties of a superfluid Fermi gas}

\author{Hiroyuki Tajima$^1$, Pieter van Wyk$^1$, Ryo Hanai$^1$, Daichi Kagamihara$^1$, Daisuke Inotani$^1$, Munekazu Horikoshi$^{2,3}$, and Yoji Ohashi$^1$}
\affiliation{$^1$Department of Physics, Keio University, 3-14-1 Hiyoshi, Yokohama 223-8522, Japan}
\affiliation{$^2$Institute for Photon Science and Technology, Graduate School of Science, The University of Tokyo, 7-3-1 Hongo, Bunkyo-ku, Tokyo 113-0033, Japan}
\affiliation{$^3$Photon Science Center, Graduate School of Engineering, The University of Tokyo, 2-11-16, Yayoi, Bunkyo-ku, Tokyo 113-8656, Japan} 
\date{\today}
\begin{abstract}
We theoretically present an economical and convenient way to study ground-state properties of a strongly interacting superfluid Fermi gas. 
Our strategy is that complicated strong-coupling calculations are used only to evaluate quantum fluctuation corrections to the chemical potential $\mu$. 
Then, without any further strong-coupling calculations, we calculate the compressibility, sound velocity, internal energy, pressure, and Tan's contact, from the calculated $\mu$ without loss of accuracy, by using exact thermodynamic identities. Using a recent precise measurement of $\mu$ in a superfluid $^6$Li Fermi gas, we show that an extended $T$-matrix approximation (ETMA) is suitable for our purpose, especially in the BCS-unitary regime, where our results indicate that many-body corrections are dominated by superfluid fluctuations.
Since precise determinations of physical quantities are not always easy in cold Fermi gas physics, our 
approach 
would greatly reduce experimental and theoretical efforts toward the understanding of ground-state properties of this strongly interacting Fermi system.
\end{abstract}
\maketitle
\par
While the tunability of various physical parameters, such as an interaction associated with a Feshbach resonance, is an advantage of ultracold Fermi gases \cite{Bloch,Giorgini,Chin}, the fact that precise measurements are not always easy (compared to the electron condensed matter systems) is a weak point of this system. This becomes more serious in examining ground-state properties of a strongly interacting superfluid Fermi gas \cite{Leggett,NSR,Engelbrecht}, because some fundamental observables, such as the spin susceptibility \cite{Sanner} and specific heat \cite{Ku}, vanish at $T=0$. 
\par
Overcoming this difficulty may also contribute to the development of other research fields, e.g., neutron-star physics. Since the recent discoveries of massive neutron stars \cite{Shapiro1,Shapiro2}, the internal structure of a neutron star has attracted much attention with renewed interest \cite{APR,Abe,NS}. Since the low density region of a neutron-star interior is expected to be similar to a strongly interacting superfluid Fermi gas at $T\sim 0 $ \cite{Carlson,Forbes}, latter atomic system may be used as a quantum simulator for the former nuclear case. 
\par
In this letter, as a possible way to resolve the above-mentioned problem existing in cold Fermi gas physics, we theoretically present a set of ground-state quantities with high accuracy and reliability, in the BCS-unitary regime of a superfluid Fermi gas. Our strategy is that we first use the recent measurement of the chemical potential $\mu$ in this regime of a superfluid $^6$Li Fermi gas \cite{Horikoshi}, to find a strong-coupling theory which can reproduce the experimental data. Then, combining this theory with {\it exact} thermodynamic identities, we evaluate several fundamental quantities, such as compressibility $\kappa_{T}$, sound velocity $v_{\rm s}$, internal energy $E$, pressure $P$, and Tan's contact $C$ \cite{tan}, from the calculated $\mu$.
An advantage of this approach is that, all the calculated quantities have the same accuracy, because calculations from $\mu$ only rely on {\it exact} thermodynamic formulae. Thus, when one of the calculated quantities ($\equiv X$) well explains highly precise experimental data, one may understand the other quantities also have the same reliability as $X$. (In this paper, $\mu$ is used as $X$.) Another advantage is that, by grouping physical quantities in this manner, strong-coupling effects on them can be summarized as quantum fluctuation corrections to $X$.
\par
We consider a two-component homogeneous superfluid Fermi gas, described by the BCS Hamiltonian in the two-component Nambu representation \cite{Schrieffer},
\begin{eqnarray}
H=\sum_{\bm p}\Psi^{\dag}_{\bm p}\left[\xi_{\bm p}\tau_{3}-\op\tau_{1}\right]\Psi_{\bm p} -U\sum_{\bm q}\rho_{+}(\bm{q})\rho_{-}(-\bm{q}).
\label{eq1}
\end{eqnarray}
In this letter, we take $k_{\rm B}=\hbar=1$, and the system volume $V$ is taken to be unity. 
In Eq. (\ref{eq1}), $\Psi_{\bm{p}}=(c_{\bm{p},\up},c_{\bm{-p},\dwn}^\dag)^T$ is the two-component Nambu field, and $\tau_{i=1,2,3}$ are the corresponding Pauli matrices. $c_{\bm{p},\sigma}$ is the annihilation operator of a Fermi atom with pseudospin $\sigma=\uparrow,\downarrow$, describing two atomic hyperfine states.
$\xi_{\bm{p}}={\bm p}^2/(2m)-\mu$ is the kinetic energy of a Fermi atom with a mass $m$, measured from the chemical potential $\mu$. $\Delta$ is the superfluid order parameter, which is taken to be real and parallel to the $\tau_1$-component, without loss of generality. $\rho_{\pm}=[\rho_1(\bm{q})\pm i\rho_2(\bm{q})]/2$ is the generalized density operator, where $\rho_1({\bm q})=\sum_{\bm p}\Psi_{\bm{p}+\bm{q}/2}^{\dag}\tau_1\Psi_{\bm{p}-\bm{q}/2}$ and $\rho_2({\bm q})=\sum_{\bm p}\Psi_{\bm{p}+\bm{q}/2}^{\dag}\tau_2\Psi_{\bm{p}-\bm{q}/2}$ physically mean amplitude and phase fluctuations of $\Delta$, respectively \cite{Ohashi2,Fukushima}. We measure the interaction strength in terms of the $s$-wave scattering length $a_s$, which is related to a bare attractive interaction $-U$ as $m/(4\pi a_s)=-U^{-1}+\sum_{\bm{p}}m/{\bm p}^2$. 
\par
\begin{figure}[t]
\begin{center}
\includegraphics[width=6.75cm]{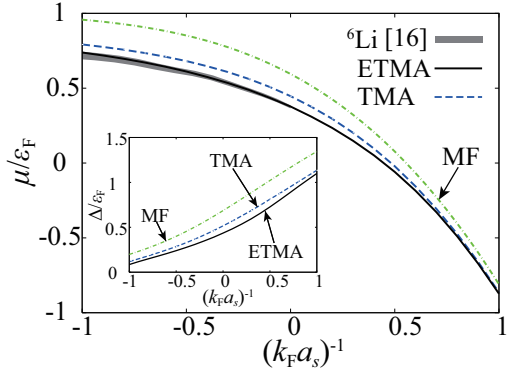}
\end{center}
\caption{(color online) Calculated chemical potential $\mu$ in ETMA, in the BCS-BEC crossover regime of a superfluid Fermi gas. 
$\varepsilon_{\rm F}$ and $k_{\rm F}$ are the Fermi energy and Fermi momentum, respectively. The gray curve shows the recent experiment on a $^6$Li superfluid Fermi gas in the BCS-unitary regime at $T/T_{\rm F}\simeq 0.06$ \cite{Horikoshi}. Following this experiment, we also set $T/T_{\rm F}=0.06$. TMA: non-selfconsistent $T$-matrix approximation. MF: BCS-Leggett theory. The inset shows the superfluid order parameter $\Delta$. 
}
\label{fig1}
\end{figure}
\par
The first step is to find a strong-coupling theory which can reproduce the recently observed chemical potential $\mu$ in a $^6$Li superfluid Fermi gas far below the superfluid phase transition temperature $T_{\rm c}$ ($T/T_{\rm F}\simeq 0.06$, where $T_{\rm F}$ is the Fermi temperature) \cite{Horikoshi}.
In this regard, Fig. \ref{fig1} shows that an extended $T$-matrix approximation (ETMA) \cite{Kashimura,Tajima,Tajima2} well explains this result, without any fitting parameters. ETMA gives the value of the Bertsch parameter \cite{Baker} as $\xi_{\rm B}=0.381$, which is also close to $\xi_{\rm B}=0.376(4)$ obtained by another experiment \cite{Ku}. We briefly note that, because of $T_{\rm c}/T_{\rm F}\sim 0.2\gg 0.06$ in the unitary regime, $\mu$ shown in Fig. \ref{fig1} is actually almost the same as the ground-state result in this region \cite{note4}. 
\par
\begin{figure}[t]
\begin{center}
\includegraphics[width=7.5cm]{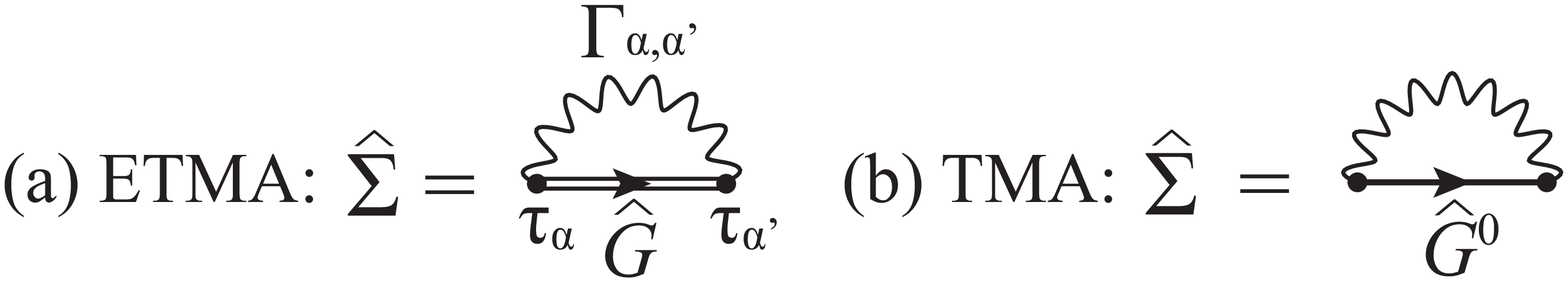}
\end{center}
\caption{(color online) Feynman diagrams describing the self-energy ${\hat \Sigma}$. (a) ETMA. (b) TMA. The double and single solid lines represent the dressed Green's function $\hat{G}$ and the bare one $\hat{G}^{0}$, respectively. The wavy line shows the particle-particle scattering matrix $\Gamma_{\alpha,\alpha'}$. The solid circles are Pauli matrices.}
\label{fig2}
\end{figure}

ETMA is characterized by a $2\times 2$-matrix self-energy $\hat{\Sigma}(p)$ in the $2\times 2$-matrix single-particle thermal Green's function $\hat{G}(p)=[\hat{G}^0(p)^{-1}-\hat{\Sigma}(p)]^{-1}$. Diagrammatically, the ETMA $\hat{\Sigma}(p)$ is given as Fig. \ref{fig2}(a) (where $\hat{G}^0(p)=[i\omega_n -\xi_{\bm{p}}\tau_3 +\Delta \tau_1]^{-1}$ is the BCS Green's function in the Nambu representation) \cite{note}. In Fig. \ref{fig2}(a), the particle-particle scattering matrix,
\begin{eqnarray}
\left(
\begin{array}{cc}
\Gamma_{-+} & \Gamma_{--} \\
\Gamma_{++} & \Gamma_{+-} \\
\end{array}
\right)
=-U
\left[
1+U
\left(
\begin{array}{cc}
\Pi_{-+} & \Pi_{--} \\
\Pi_{++} & \Pi_{+-} \\
\end{array}
\right)
\right]^{-1},
\label{eq2c}
\end{eqnarray}
describes superfluid fluctuations, where 
\begin{equation}
\Pi_{\alpha,\alpha'}(q)=T\sum_{p}{\rm Tr}\left[\tau_\alpha\hat{G}^0(p+q)\tau_{\alpha'}\hat{G}^0(p)\right]
\label{eq2b}
\end{equation}
is a pair-correlation function. The expression for the ETMA self-energy is given by
\begin{equation}
\hat{\Sigma}(p)
=-T\sum_{q}\sum_{\alpha,\alpha'=\pm}
\Gamma_{\alpha,\alpha'}(q)
\tau_{\alpha}\hat{G}(p+q)\tau_{\alpha'}.
\label{eq2}
\end{equation}
The ETMA chemical potential $\mu$ in Fig. \ref{fig1} and the superfluid order parameter $\Delta$ shown in the inset in Fig. \ref{fig1} are self-consistently determined by numerically solving the number equation, $n=T\sum_{p}{\rm Tr}[\tau_3 \hat{G}(p)]$
, together with the gap equation,
\begin{equation}
1=-{4\pi a_s \over m}
\sum_{\bm p}
\left[
{1 \over 2E_{\bm p}}{\rm tanh}\frac{E_{\bm{p}}}{2T}-\frac{m}{{\bm p}^2}
\right],
\label{eq4}
\end{equation}
where $E_{\bm p}=\sqrt{\xi_{\bm p}^2+\Delta^2}$ is the Bogoliubov dispersion \cite{Tajima2}. 
\par
\par
Although it is believed that the BCS-Leggett theory \cite{Leggett} can qualitatively describe BCS-BEC crossover physics at $T=0$, Fig. \ref{fig1} shows that it quantitatively overestimates the magnitude of $\mu$. Since thermal fluctuations are suppressed far below $T_{\rm c}$, the difference between the ETMA result and this mean-field result seen in Fig. \ref{fig1} comes from quantum fluctuations existing even at $T=0$. Figure \ref{fig1} also shows that the inclusion of many-body corrections to $\mu$ is insufficient in the non-selfconsistent $T$-matrix approximation (TMA) \cite{Palestini,Pieri2,Watanabe}. Here, the TMA self-energy is given by replacing the dressed Green's function ${\hat G}$ in Eq. (\ref{eq2}) with the bare one ${\hat G}^0$ (see also Fig. \ref{fig2}(b)). The (strong-coupling) Luttinger-Ward approach (LW) \cite{Haussmann2}, which is given by replacing all the bare Green's functions ${\hat G}^0$ in the pair-correlation function in Eq. (\ref{eq2b}) by the dressed ones ${\hat G}$, gives $\mu(T=0)/\varepsilon_{\rm F}=0.36$ in the unitary limit (where $\varepsilon_{\rm F}$ is the Fermi energy), which is somehow smaller than the experimental value ($\mu/\varepsilon_{\rm F}=0.38$) \cite{Horikoshi,Ku}, indicating slight overestimation of quantum fluctuations.
\par
To see the background physics of strong-coupling corrections to $\mu$, it is convenient to approximately treat the particle-particle scattering matrix $\Gamma_{\alpha\alpha'}$ in Eq. (\ref{eq2c}) as a constant $\Gamma_{\rm eff}~(<0)$, and extract the $\tau_3$-component from the self-energy ($\equiv {\hat \Sigma}_3$), which has the form $\hat{\Sigma}_3=\Gamma_{\rm eff}(n/2)\tau_3$ in ETMA. When we only include this effect, the resulting $\mu$ shifts from the BCS-Leggett result ($\mu_{\rm MF}$) as $\mu=\mu_{\rm MF}-|\Gamma_{\rm eff}|n/2 <\mu_{\rm MF}$, which qualitatively explains the reason for the smaller $\mu$ in ETMA compared to the BCS-Leggett result. A similar correction is also obtained in TMA, where the number density $n$ in the correction term $\delta\mu=-|\Gamma_{\rm eff}|n/2$ is replaced by the mean-field number density $n_0=T\sum_{p}{\rm Tr}[\tau_3 \hat{G}^0(p)]$, reflecting the difference between ETMA and TMA self-energies shown in Fig. \ref{fig2}. Since $n_0$ decreases from $n$ with increasing the interaction strength in the BCS-unitary regime \cite{NSR}, the TMA correction becomes smaller than the ETMA case, as shown in Fig. \ref{fig1}. We note that, although the correction $\delta\mu=-|\Gamma_{\rm eff}|n/2$ looks similar to the ordinary Hartree shift $E_{\rm Hartree}=-Un/2$, $E_{\rm Hartree}$ actually vanishes in ETMA, as well as in TMA, because of the vanishing bare interaction $(U\to +0)$ in these renormalized theories with an infinitely large energy cutoff. Instead, $\delta\mu(T=0) $ comes from superfluid fluctuations \cite{Kinnunen,Schirotzek,Sagi} existing even at $T=0$.
\par
\begin{figure}[t]
\begin{center}
\includegraphics[width=6.75cm]{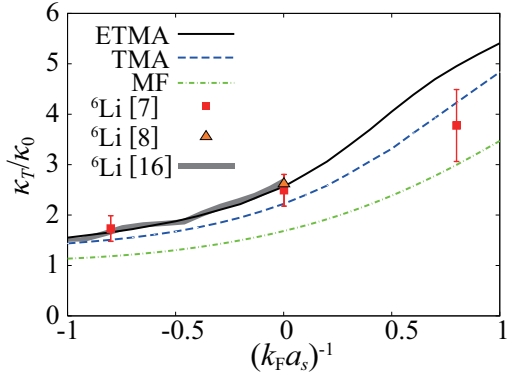}
\end{center}
\caption{(color online) Calculated isothermal compressibility $\kappa_T$ in the BCS-BEC crossover region at $T/T_{\rm F}=0.06$. $\kappa_0=3/(2n\varepsilon_{\rm F})$ is the compressibility in a free Fermi gas at $T=0$. In this figure, we use the same line styles as those in Fig. \ref{fig1}.
}
\label{fig3}
\end{figure}
\par
We now employ ETMA to examine other ground-state quantities in the BCS-BEC crossover region. As far as we use ETMA only for the purpose of the evaluation of $\mu$ appearing in an exact thermodynamic expression for a physical quantity $X$, the calculated $X$ should still have the same accuracy as the ETMA $\mu$ in Fig. \ref{fig1}. 
\par
The first non-vanishing example is the isothermal compressibility $\kappa_T$. This can be obtained from $\mu$ via the thermodynamic identity,
\begin{eqnarray}
\kappa_{T}
&=&{1 \over n^2}\left(\frac{\partial n}{\partial \mu}\right)_T.
\label{eqk}
\end{eqnarray}
Figure \ref{fig3} shows $\kappa_T(T/T_{\rm F}=0.06)$ obtained by numerically evaluating the derivative in Eq. (\ref{eqk}) by considering two cases with slightly different densities in ETMA. In the BCS-unitary regime, we see that the calculated $\kappa_T$ agrees well with the experiment on a $^6$Li Fermi gas \cite{Horikoshi}, as well as other two experiments on $^6$Li Fermi gases \cite{Sanner,Ku}. On the other hand, the ETMA result deviates from the observed $\kappa_T$ in the BEC regime when $(k_{\rm F}a_s)^{-1}\simeq 0.8$ \cite{Ku}, which we will comment on later.
\par
The larger $\kappa_T$ in ETMA than the mean-field result in Fig. \ref{fig3} indicates the importance of the Stoner enhancement. 
When we use Eq. (\ref{eqk}) to calculate $\kappa_T$ using the ETMA Green's function ${\hat G}$, the Ward identity \cite{Mahan} is automatically satisfied, which guarantees consistency between the self-energy and the three-point vertex for $\kappa_T$. In ETMA, this three-point vertex consists of RPA (random-phase approximation) type infinite series of bubble diagrams. The resulting ETMA compressibility symbolically has the form $\kappa_T\sim\kappa_T^{\rm MF}/[1-W\kappa_T^{\rm MF}]$ (where $W$ is a positive constant). The Stoner factor, $1-W\kappa_T^{\rm MF}~(<1)$, enhances $\kappa_T$ compared to the mean-field value $\kappa_T^{\rm MF}$, as seen in Fig. \ref{fig3}. In TMA, on the other hand, the consistent three-point vertex to the TMA self-energy is given by truncating the RPA series up to $O(W)$, leading to $\kappa_T\sim \kappa_T^{\rm MF}[1+W\kappa_T^{\rm MF}]$. Thus, although the Stoner enhancement is partially included in TMA, the TMA compressibility is smaller than the ETMA case, as shown in Fig. \ref{fig3}.
\begin{figure}[t]
\begin{center}
\includegraphics[width=6.75cm]{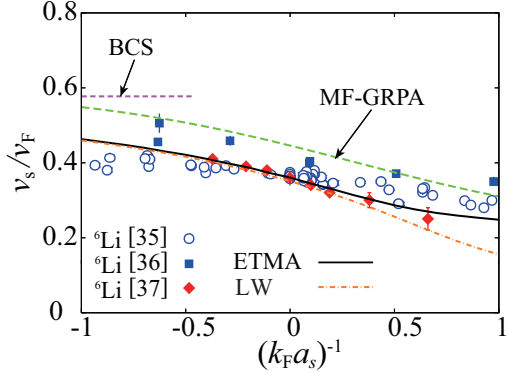}
\end{center}
\caption{(color online) Calculated sound velocity $v_{\rm s}$ at $T=0$, normalized by the Fermi velocity $v_{\rm F}$. BCS: the weak-coupling BCS result, $v_{\rm s}=v_{\rm F}/\sqrt{3}$ \cite{Engelbrecht,Ohashi2}. MF-GRPA: combined BCS-Leggett theory with GRPA. LW: Luttinger-Ward approach \cite{Haussmann2}. In calculating $v_{\rm s}$ from Eq. (\ref{eq3}), we have approximately used $\kappa_T$ in Fig. \ref{fig3} for the compressibility at $T=0$.}
\label{fig4}
\end{figure}
\par
Noting that the adiabatic compressibility $\kappa_S$ coincides with $\kappa_T$ at $T=0$ because of the vanishing entropy $S(T=0)$, we can evaluate the sound velocity $v_{\rm s}(T=0)$ with the same accuracy as $\mu$ and $\kappa_T$ from
\begin{equation}
v_{\rm s}(T=0)={1 \over\sqrt{n m\kappa_S}}={1 \over\sqrt{n m\kappa_T}}.
\label{eq3}
\end{equation}
Since the calculated $v_{\rm s}$ is supported by the experiment on $\mu$ in the BCS-unitary regime \cite{Horikoshi}, it would give a constraint to experiments in this region. Figure \ref{fig4} shows that, among the three experiments \cite{Joseph,Weimer,Vale}, the observed $v_{\rm s}$ by the Bragg spectroscopy \cite{Vale} is in good agreement with our result. Figure \ref{fig4} also shows that, compared to the result by the combined mean-field theory with the generalized random-phase approximation (MF-GRPA) \cite{Fukushima}, $v_{\rm s}$ in ETMA is away from the weak-coupling BCS result even at $(k_{\rm F}a_s)^{-1}=-1$, indicating the importance of strong-coupling corrections even there. Indeed, ETMA sound velocity agrees with $v_{\rm s}$ obtained by LW \cite{Haussmann2} in the BCS regime (see Fig. \ref{fig4}). The difference between ETMA and LW seen in the BEC side might come from the different treatments of collective modes between the two theories \cite{note5}.  
\par
However, our approach has room for improvement in the BEC regime. In this regime, the sound mode is described by the Bogoliubov phonon in a molecular BEC with a repulsive interaction $U_{\rm M}=4\pi a_{\rm M}/(2m)$. Since ETMA overestimates the molecular scattering length as $a_{\rm M}=2a_s$ in this regime (Note that the correct value equals $a_{\rm M}=0.6a_s$ \cite{Petrov}.), ETMA would also overestimate $v_{\rm s}~(\propto\sqrt{U_{\rm M}})$ there. Other quantities in ETMA would also be affected by this overestimation in the BEC region. The discrepancy between the ETMA compressibility and the experiment \cite{Ku} in this regime shown in Fig. \ref{fig3} also implies the necessity of a strong-coupling theory beyond the current ETMA \cite{note3}. To see to what extent our combined ETMA approach with exact thermodynamic identities works in the BEC regime, highly accurate experimental data for $\mu$ in this regime would be helpful. However, one should note that our approach using exact thermodynamic identities is not restricted to the validity of ETMA. That is, once one can replace ETMA by a more sophisticated theory which quantitatively well describes $\mu$ in the BEC regime, our approach using exact thermodynamic identities can again evaluate other physical quantities in the BEC regime with high accuracy as $\mu$, as in the case of the BCS side. 
\par
\begin{figure}[t]
\begin{center}
\includegraphics[width=6.75cm]{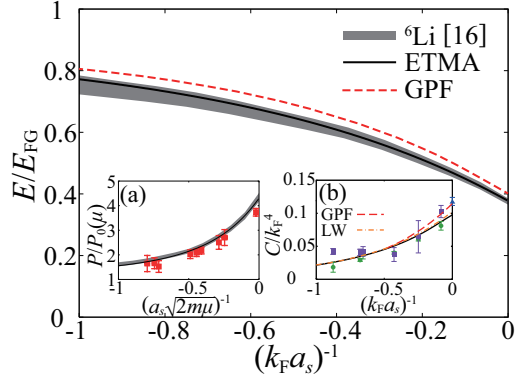}
\end{center}
\caption{(color online) Calculated ground-state energy $E$ of a superfluid Fermi gas in the BCS-unitary region. $E_{\rm FG}=(3/5)n\varepsilon_{\rm F}$ is the ground-state energy of a free Fermi gas. The solid and dashed lines show results of ETMA and Gaussian pair fluctuation theory (GPF), respectively. The insets (a) and (b) show, respectively, the ground-state pressure $P$ and Tan's contact $C$. In the inset (a), squares represent an experiment on a $^6$Li Fermi gas \cite{ENS}. $P_0(\mu)=(2(2m)^{3/2}/(15\pi^2))\mu^{5/2}$ is the pressure of a free Fermi gas at $T=0$. In the inset (b), filled circles and squares are experimental data on a $^{40}$K Fermi gas \cite{JinC}, and filled triangles are experimental data on a $^6$Li Fermi gas \cite{ENS}.}
\label{fig5}
\end{figure}
\par
As shown in Fig. \ref{fig5}, the ground-state energy $E$ can also be obtained from $\mu$, via the differential equation \cite{note2}, 
\begin{equation}
\frac{\mu}{\varepsilon_{\rm F}}=\frac{E}{E_{\rm FG}}-\frac{(k_{\rm F}a_s)^{-1}}{5}\frac{d(E/E_{\rm FG})}{d(k_{\rm F}a_s)^{-1}},
\label{eq5}
\end{equation} 
where $E_{\rm FG}=(3/5)n\varepsilon_{\rm F}$ is the ground-state energy of a free Fermi gas. One can then obtain the pressure $P(T=0)=-E+\mu n$ shown in the inset (a) in Fig. \ref{fig5}. ETMA also agrees with the ENS experiment \cite{ENS}. We briefly note that the Gaussian pair fluctuation theory (GPF) \cite{Hu} slightly overestimates the internal energy $E$ (see Fig. \ref{fig5}), which is because GPF underestimates many-body corrections to $\mu$ compared to ETMA.  
\par
The accuracy of the calculated internal energy in Fig. \ref{fig5} is supported by the experiment on $\mu$ \cite{Horikoshi}. In addition to this, the correctness of this result can also be checked by further calculating the Tan's contact from $C=-4\pi m(\partial E/\partial a_s^{-1})$. As shown in the inset (b) in Fig. \ref{fig5}, the calculated $C$ agrees well with the recent experiments \cite{JinC,ENS}, LW \cite{Haussmann3}, as well as GPF \cite{Hu2}. Furthermore, at the unitarity, ETMA result ($C/k_{\rm F}^4=0.098$) also agrees with the experiment on a $^6$Li Fermi gas ($C/k_{\rm F}^4=0.107(3)$) \cite{Hoinka2}, a quantum Monte-Carlo (QMC) result ($C/k_{\rm F}^4=0.0996(34)$) \cite{QMC}, as well as fixed-node diffusion Monte-Carlo (FNDMC) calculation ($C/k_{\rm F}^4=0.1147(3)$) \cite{FNDMC}. 
\par
Although a strongly interacting superfluid Fermi gas at $T\ll T_{\rm c}$ is a candidate for a quantum simulator to study the neutron-star interior in the low density region, one should note that the effective range $r_{\rm eff}$ is different between the two. While $r_{\rm eff}$ can be safely ignored in the former atomic system, it cannot be ignored in the latter, because the value $r_{\rm eff}=2.7~{\rm fm}$ becomes comparable to $k_{\rm F}^{-1}$ even in the relatively low density region. Since it is difficult to tune $r_{\rm eff}$ in the current experimental stage of cold atom physics, we need to make up for this difference theoretically, when we explore the neutron-star interior with the help of cold Fermi gas physics. Our results indicate that ETMA may be a good starting point for this purpose. 
\par
To summarize, we have discussed ground-state quantities in a strongly interacting superfluid Fermi gas. Instead of independently evaluating them, we first confirmed that ETMA can well reproduce the recently observed chemical potential $\mu$ in a $^6$Li superfluid Fermi gas \cite{Horikoshi}. Then, combining ETMA with exact thermodynamic identities, we evaluated the other quantities in this regime from the calculated $\mu$, without loss of accuracy. To confirm the validity of this approach, we showed that some of our results agree with recent experiments (that are different from the experiment on $\mu$). We also pointed out that strong-couping effects on these quantities in the ground-state may be summarized as quantum fluctuation corrections to $\mu$. 
\par
We thank C. J. Vale for providing us his experimental data, as well as B. Frank and W. Zwerger for sharing their updated numerical results of those in Ref.\cite{Haussmann2}. We also thank T. Hatsuda and M. Matsumoto for useful discussions. H.T. and R.H. were supported by a Grant-in-Aid for JSPS fellows. This work was supported by KiPAS project in Keio University, as well as Grant-in-aid for Scientific Research from MEXT and JSPS in Japan (No.JP16K17773, No.JP24105006, No.JP23684033, No.JP15H00840, No.JP15K00178, No.JP16K05503). 


\begin{thebibliography}{9}
\bibitem{note4} As shown in Ref. \cite{Tajima2}. the temperature dependence of $\mu$ is very weak below $T_{\rm c}$, especially when $T\lesssim 0.06T_{\rm F}$. We also note that $T_{\rm c}$ at $(k_{\rm F}a_s)^{-1}=-1$ is still higher than $0.06T_{\rm F}$. Thus, Fig. \ref{fig1} shows the case below $T_{\rm c}$.
\bibitem{note} To save space, we use the abbreviated notations, $p=({\bm p},i\omega_n)$ and $q=({\bm q},i\nu_n)$, where $\omega_n$ and $\nu_n$ are fermion and boson Matsubara frequencies, respectively.
\bibitem{note5} Since the LW approach \cite{Haussmann2} gives gapped collective excitations, a modified coupling constant in the gap equation is employed to avoid this singularity. This manipulation is not necessary for ETMA, because the gap equation (\ref{eq4}) is already consistent with the condition that $\hat{\Gamma}$ has a pole at ${\bm q}=\nu_n=0$.
\bibitem{note3} However, since Ref.\cite{Sanner} measures $\kappa_T$ in the BEC side at $T/T_{\rm F}\simeq 0.19$ (which is close to $T_{\rm c}$), temperature effects may also be an origin of this discrepancy.
\end{thebibliography}
\end{document}